# Creative elements: network-based predictions of active centres in proteins, cellular and social networks

**Peter Csermely**[1]

Department of Medical Chemistry, Semmelweis University, P O Box 260., H-1444 Budapest 8, Hungary

**Active centres and hot spots of proteins have a paramount importance in enzyme action, protein complex formation and drug design. Recently a number of publications successfully applied the analysis of residue networks to predict active centres in proteins. Most real-world networks show a number of properties, such as small-worldness or scale-free degree distribution, which are rather general features of networks, from molecules to society at large. Using analogy I propose that existing findings and methodology already enable us to detect active centres in cells, and can be expanded to social networks and ecosystems. Members of these active centres are termed here as 'creative elements' of their respective networks, which may help them to survive unprecedented, novel challenges, and play a key role in the development, survival and evolvability of complex systems.**

## Introduction: various approaches to determine key elements of protein structures

Active centres of proteins have been first identified by their function, e.g. by their participation in enzyme catalysis and substrate binding. The well-known 'lock and key model' whereby the conformation of the active centre of a protein (the lock) was only matched by its unique ligand(s) (the key), was proposed by Emil Fischer in the 19[th] century [1] and was used to describe the mechanism of enzyme action for a long time. It wasn't until fifty years ago that Daniel Koshland challenged this view and proposed the 'induced fit mechanism' whereby the binding of the correct ligand(s) led to the protein changing its shape and adopting its active conformation, a concept which became a centrepiece of our biochemical understanding of enzyme function [2]. More recent developments emphasize the importance of atomic vibrations of the protein structure, and the partner-driven selection of the binding-compatible conformation from an ensemble of alternating conformations of the original protein. This mechanism was first outlined in 1964 by Bruno Straub, who called it 'fluctuation fit' and, more recently by Nussinov and colleagues, as a modern day interpretation called the 'pre-existing equilibrium/conformational selection model' [3,4]. Another way to achieve 'fluctuation' is structural disorder in a large number of protein-complexes termed as 'fuzziness' by Peter Tompa [5]. Here it is worth emphasizing that from the point of the general elements of the binding mechanism of a protein, the binding partner can be not only a substrate, but also any other small ligand, drug or even a macromolecule, such as another protein, DNA or RNA. An increase in the size and binding strength of the partner was proposed to shift the preferred binding mechanism from the fluctuation fit towards the induced fit [6]. Due to the concentration of binding free energy to a small number of critical amino acid residues of protein binding surfaces, these residues were termed as 'hot spots' ([7], for further elaboration on the terminology of active centres and hot spots, see Box 1). Hot spots often cluster to densely-packed 'hot regions' [8]. Active centres are thus not just binding pockets acting as baits waiting for their prey, but must also have a special position in the protein structure to trigger a set of concerted conformational changes.

How can we predict active centres and hot spots? During the past decades several methods have been developed, which are able to predict active centres and their key residues with high accuracy such as SCOTCH, MAPPIS, KFC-server and CS-Map (for many more freely available programs, see Tables 6. and 7. of ref. [9]). These methods use the evolutionary conservation of physico-chemical properties, energy-optimization, neural networks or machine learning. However, most of these approaches are dependent on structural and energy-based information, which is local, or need the extensive comparison of evolutionary variants [9-12]. The special position of active centres in the overall structure of the hosting protein may allow the use of global structural determinants to identify additional discriminatory features helping their prediction from a single protein structure.

*Corresponding author:* Csermely, P. (csermely@puskin.sote.hu)

> **Box 1. Active centres and hot spots**
>
> The terminology of important protein residues and regions is rather complex due to the various approaches identifying these critical segments of protein structure. Here I outline the various approaches and the related terms.
> 1. **Functional terms: active centres and binding sites.** In the traditional sense we call a protein segment an active centre, if it plays a key role in the catalytic action of the enzyme function displayed by the respective protein. Similarly, amino acid side chains located at the binding interface are called as the binding site. However, in both cases the protein segments, which functionally participate in either the catalytic or the binding action are much broader than those identified by the traditional terms of the active centre or binding site.
> 2. **Energy terms: hot spots and hot regions.** A small number of critical amino acid residues of protein binding surfaces can be discriminated by an extraordinarily high binding free energy, and have been termed as 'hot spots' [5]. Hot spots often cluster to densely-packed 'hot regions' [6].
> 3. **Structural terms: central residues, active centres and creative elements.** Description of protein structures as amino acid networks allowed the identification of central network residues. These residues are often clustered and form active centres in the structural, network sense. With the exception of the Introduction, throughout this paper I use the term 'active centre' in this novel, network sense. A highly specific subset of these central residues has been described in this paper as creative elements. For the discriminatory features of creative elements see the main text.
>
> The terminologies defined by the three approaches overlap each other. Hot spots and hot regions are parts of binding sites. Central residues often contain catalytic residues (active centres in the traditional sense), as well as segments of binding sites including hot spots. However, central amino acid network residues including creative elements often go beyond the traditionally identified key segments and highlight novel critical residues governing conformational changes. The above terms are often compared to and extended by mutational studies as well as by evolutionary analysis. As expected, mutations inducing more damage in protein function as well as evolutionary conserved residues are often (but by far not always) overlapping with the active centres, binding sites, hot spots and central residues.

In the first section of this paper I will summarize our knowledge on protein structure networks (residue networks), which will be followed by the description of network-based prediction of active centres. I will show that amino acids of active centres have a number of discriminatory features at their own, network element level. Since (1) all these discriminatory features are independent from the functional identification of active centres using the properties of the whole network together (thus a system one level higher in the hierarchical organization of nature), and (2) they identify a broader set of crucially important amino acids than the amino acids of traditional active centres (see Box 1), and (3) correspond to the network behaviour of creative persons, I will propose that it is worth to discriminate key residues of protein structure networks and call them as 'creative elements'. In the concluding parts of the paper I will give a number of examples to show that in other networks, such as in protein–protein interaction networks, signalling networks, social networks and ecosystems we can also identify highly similar creative elements, thus their existence seems to be a general feature of evolving systems. The examples of the paper will show that creative elements help the survival of unprecedented, novel challenges, and play a key role in the development and evolvability of complex systems.

**Protein structure networks**

The role of a certain amino acid at a certain position in the protein structure may be assessed by the network approach. Networks help us to understand complex system behaviour by reducing the system to a set of interacting elements, which are bound together by links [13-15]. In protein structure networks (also known as residue networks) the interacting elements are the amino acids of protein molecules, while the links represent their neighbouring position in space if the inter-element distance is below a cut-off of usually between 0.45 and 0.85 nm. Protein structure networks may use weights instead of the cut-off, may be restricted to hydrogen bonds only, and may also define each individual atom of the protein structure as an element [16,17]. Proteins are small-worlds (see Glossary). In the small-worlds of protein structure networks any two elements are connected to each other via only a few other elements represented by amino acids. Small-worldness determines folding probability (proteins with denser protein structure networks fold easier), and increases during the folding process as the protein structure becomes more and more compact [18].

Protein structure networks contain the constraints of the protein backbone only as indirect information. The neglect of the backbone-related confinement of potential amino acid motions doesn't cause any problems, if we restrict our analysis to the topology of these networks, and do not want to examine their dynamics. However, for a more complete understanding of protein dynamics different methods are also needed. The majority of the elastic network model uses the atomic coordinates of the alpha carbon atoms and a harmonic potential to account for the pair-wise interactions between them. Other elastic network representations include all atoms, forming a spring network [19-23]. As I will show later, these elastic network models may accommodate the recent knowledge on atomic vibrations in the protein structure, and allow a better understanding of the conformational selection process.

**Network-based prediction of active centres in proteins**
How can we find discriminatory features of active centres by analyzing the topology of protein structure networks? Due to the constraints imposed by the protein structure, local topological extremities do not seem to answer this question. As an example of this, 'stars' or 'mega-hubs' (i.e. elements with an extremely large number of neighbours) cannot be observed in protein structure networks because the surface area and binding properties of a single amino acid side chain do not allow the continuous binding of a large number of neighbours due to steric hindrance. The 'trick' of partner-change (used for the expansion of partners in other networks, like in protein–protein interaction networks or in human relationships) is hindered by the protein backbone [16,17]. If searching for discriminatory features of active centres, we should consider the complexity of the whole network.

Centrality is a key measure of long-range network topology. According to an often-used version of the multitude of centrality definitions called 'betweenness centrality', an element is central, if it is needed for a large number of shortest paths, where the term 'shortest path' means the shortest possible route between two elements of the network [13-15]. Ruth Nussinov and co-workers pruned the protein structure networks of seven large protein families, removing those segments which did not affect the average path-length greatly [24]. With this method they constructed a 'network-skeleton' containing only those side chains, which were central enough to play a major role in the information-flow (conformational relaxation) of the whole protein. Indeed, they found that these 'conserved interconnectivity determinants' were key elements of communication between the allosteric site(s) and the active sites (e.g. catalytic sites). It is of particular interest that in the case of the HIV-1 protease the remaining residues outside the active sites were sites whose mutations led to drug resistance [24]. Central residues with small average shortest path lengths were found to coincide with the catalytic site or ligand binding site(s). This, together with surface accessibility, proved to be a good predictor of active sites in 70% of 178 protein chains. Interestingly, not all active site residues had a high degree, i.e. a large number of neighbours in the network [25]. The above studies showed that active sites are, indeed, preferentially centred within the protein structure network. The catalytic centre of the ribosomal RNA also occupies a central position in the rRNA nucleotide network, and an additional central nucleotide, A2439 lies in the middle of the information-flow (conformational changes) of the rRNA molecule, which extends the above statements to ribozymes [26].

However, the residue-networks mentioned so far are static, and reflect only a single conformation of the entire conformational ensemble of the protein. Changes in protein conformation may rearrange the centrality of individual residues. Indeed, shifts in residue centrality ranks were observed, when the active and inactive conformations of hemoglobin and nitrogen regulatory protein C were compared [24]. Moreover, network centrality, when applied alone may identify additional key residues besides active centres, such as allosteric sites, hinge-elements, etc. A better prediction thus requires additional information, which may come from protein dynamics. Indeed, central amino acids have a more restricted motion [27] raising the possibility that elastic network models may reveal additional discriminatory features of active centres.

Using the structural perturbations of the elastic network model a set of sparsely connected, highly conserved residues were identified, that are key elements for the transmission of allosteric signals in three nanomachines: DNA polymerase, myosin and the GroEL chaperonin [28]. The combination of the elastic network model with information diffusion revealed that active centres are endowed with fast and precise communication [29]. A perturbation study of the conformational ensemble of dihydrofolate reductase [30] showed that the binding sites have a greater impact on the cooperation of residue pairs than any other segments of the protein. These findings suggest that active centres are not only structurally central in protein structure networks, but also have a central position to affect protein dynamics.

Recently the elastic network model has been extended by adding anharmonic, nonlinear terms and by taking into account the fact that the energy of the surface amino acids is dissipated by the surrounding water. This approach identified active centres as special, energy-preserving protein segments in 833 enzymes. The active centres collecting and harbouring long-lived, localized vibrations, called 'discrete breathers' were located on the stiffest parts of the proteins, and had many neighbours which were not preferentially connected to each other [31,32]. The uniquely high local energy of active sites is in agreement with the preferential local unfolding of these sites [33], as well as with their high local 'frustration', i.e. low contribution to the stabilization energy of the protein [34].

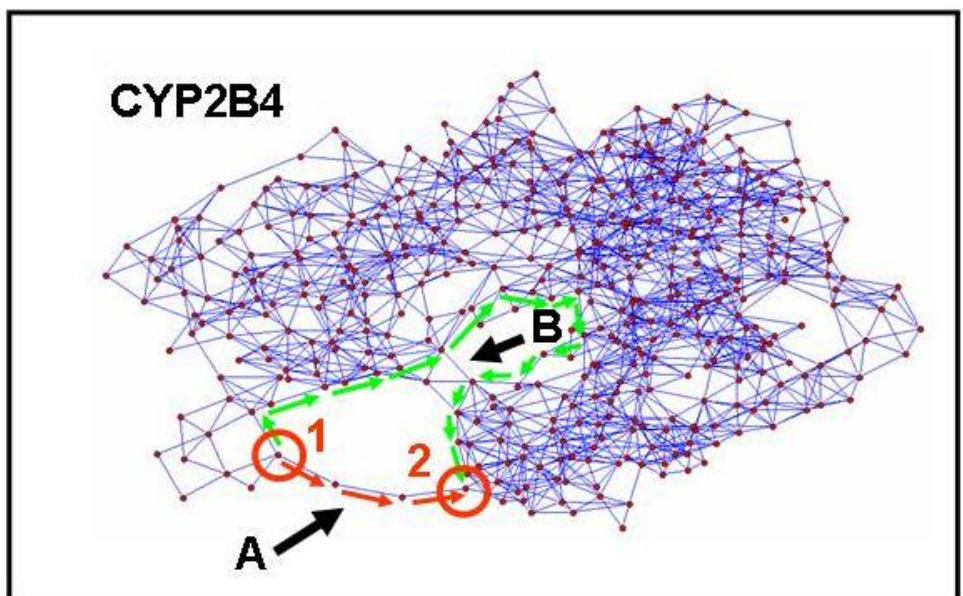

**Figure 1. Active centres of protein structural networks.** The figure illustrates the key role of active centres of protein structural networks in the communication of the protein molecule using the example of the hydrogen-bond network (made by the HBPlus program of ref. [60] and the Pajek program with the Kamada-Kawai algorithm) of the open conformation of the cytochrome P450 protein, CYP2B4 (pdb file: 1po5, leaving out the hydrogen bonds of the heme and water molecules). The black arrows point to two active centres of the hydrogen bond structural network (marked with "A" and "B"). These active centres play a key role in the communication of the network as it is shown by the red arrows, which highlight a 3 steps long communication path between sites "1" and "2" (marked with the red circles), whereas in the absence of active centres "A" and "B" the alternative communication pathways (exemplified by the one marked with the green arrows) would provide a much longer (in this particular case 21 steps long) communication path. The active centre "A" of the network contains Ile114 and corresponds to the substrate recognition site, while the active centre "B", containing Arg98 and Arg133 forms the heme-binding site illustrating the overlap of structurally important segments of protein networks with functionally important sites of the protein molecule.

The network analysis showed that active centres (1) occupy a central position in protein structure networks; (2) most of the time, but not always, are hubs, i.e. have many neighbours; (3) give non-redundant, unique connections in their neighbourhood; (4) integrate the communication of the entire network; (5) are individual, and do not take part in the dissipative motions of 'ordinary' residues and (6) collect and accommodate most of the energy of the whole network (Fig. 1). Let me note here, that the above features are not only characteristic to the bona fide active sites of enzymes, but also to the binding sites of ligands, drugs, proteins, DNA or RNA. In summary, active centres are different: they have unique properties (being stiff when the rest of the protein is flexible, being 'frustrated', etc.) and they influence the communication of all other network elements while maintaining their individuality. In analogous terms, the above summary may well sound like the characterization of a mastermind, broker, innovator or network entrepreneur. Indeed, as we proposed in preliminary forms before [35,36], and as I will show in the next section, elements with a similar network position and features may be found not only in networks of proteins, signal transduction, and neurons but also in social networks and ecosystems.

### Active centres, creative elements of cells, ecosystems and social networks

Networks have several properties, which are typical to most of them. The already mentioned small-worldness of protein structure networks is an example of one of these generally valid properties of most complex systems ensuring the fast and undistorted propagation of information in the whole network. Many real-world networks have a scale-free degree distribution, i.e. a highly uneven distribution of connections resulting in the presence of hubs and elements with many more neighbours than the average [13,14,37]. Most networks are modular, and have a well-developed hierarchical structure of overlapping groups. The module-hierarchy often leads to a fractal-like structure, where the individual structures of the different hierarchical levels resemble each other [35,38]. The hierarchical modularity allows for easy navigation in the network and (together with the presence of hubs) the filtering out of unwanted information and noise. Moreover, several pieces of evidence suggest that networks are generally stabilized by low-affinity, low-intensity, weak links [15,39].

The above list of general network properties (which is far from being complete) prompts me to think that the concept of active centres can be extended to networks other than protein structures. In protein–protein interaction networks a good analogue of an active centre is a date-hub, the definition of which in this context is a protein with one or two interaction surfaces, which forms complexes with different subsets of its partners at different times and at different cellular locations. Date-hubs are enriched for intrinsic disorder and structural flexibility which makes them different from other proteins, which are structurally more defined [40]. Date-hubs are preferentially located in the overlaps of multiple modules, thus their connections are non-redundant and unique [35]. As an example, molecular chaperones have an inter-modular localization and are amongst those date-hubs that constitute the true central coordinators of the cellular network [41,42]. Several chaperones are also called stress proteins or heat shock proteins, and their centrality increases in protein–protein interaction networks after stress [41]. In other words: chaperones become key integrators when the cell experiences an unexpected situation, i.e. a stress. As an additional example, central elements of signal transduction networks, exemplified by the phosphatidylinostiol-3-kinase, the AKT/PKB-kinase or the insulin-receptor substrate (IRS) families, have been termed as 'critical nodes' by Ronald C. Kahn and co-workers [43]. Critical nodes often have many isoforms, which shows the importance of the need for 'back-ups' for active centres as well as the need to extend the variability of these key elements of signal transduction further. It is of significant interest, how the signalling mediators of various membranes [44] can be included to this picture.

Going several levels of integration higher to the level of mammalian networks, top predators act as couplers of distinct and dissimilar energy channels, and by integrating the ecosystem-network increase its stability [45]. Dolphins occupying inter-modular positions in dolphin-communities were shown to act as brokers of social cohesion for the whole group [46]. In social networks the archetype of the above unique, inter-modular element is the 'stranger' described by one of the forefathers of sociology, George Simmel a hundred years ago [47]. The stranger is different from anyone else. The stranger belongs to all groups, but at the same time does not belong to any of them. A later, well-known example came from Ronald S. Burt [48], who proved that innovators and successful managers occupy 'structural holes', which are exactly the non-redundant, centrally connecting positions of the active centres in protein structure networks. People bridging structural holes have 'weak links', e.g. they often change their contacts [48]. Malcolm Gladwell describes several 'active centre figures' in his best-seller book, "Tipping point" [49]. These 'connectors' (including the famous Boston citizen of the American history, Paul Revere, who alarmed his fellows during the "Midnight Ride" to combat the coming danger at the beginning of the American Revolution) are interested in a large number of dissimilar persons and information. This wide and unbiased interest propels these boundary spanning individuals to an integrative, central position in the social and information networks. Such a person can also be imported from outside. For example, consultants typically span otherwise isolating intra-organization boundaries [50].

Going one level higher in the hierarchy again, having a central position also offers a great advantage to groups. As an example of this, biotech companies with diverse portfolios of well connected-collaborators were found to have the fastest access to novel information, and directed the evolution of the field. This was only possible in the long run if most of these connections were transient [51]. Their transient, far-reaching, exploratory contact structure helps performance only in those cases when the tasks are novel (e.g. those emerging in uncertain environments or in crisis) and require creative thinking to solve. Conversely, if the task is one that is typical, and the expertise that is already present within the group is enough to solve it, the maintenance of exploratory contacts is costly and hinders performance [52,53].

The above analogies enrich the characteristics of active centres further. Active centres of networks at higher levels than proteins are not only central elements having a unique set of properties and integrating the communication of the entire network, but they also perform a partially random sampling of the whole network, and connect distant modules. Active centres (especially those of less constrained networks than proteins) have transient, weak links leading to important positions (often hubs) in the network, and become especially important when the whole system experiences an atypical situation requiring a novel, creative solution. Due to this critical property of active centres, I propose the term 'creative elements' for the network elements participating in active centres (Fig. 2).

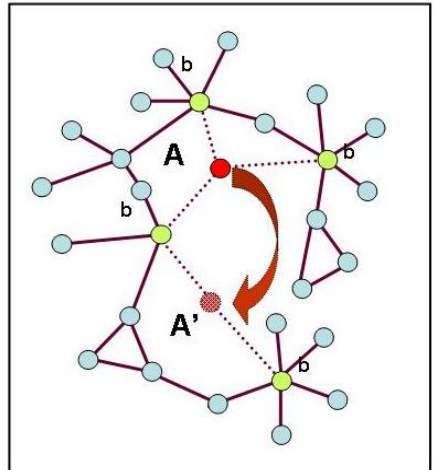

**Figure 2. Creative elements.** The figure illustrates the integrative network position of creative elements integrating network communication, performing a partially random sampling of the network, and connecting distant network modules. Creative elements often have transient, weak links leading to hubs. Creative elements (A and A') and hubs (b) are marked with red and green circles, respectively. Solid lines denote strong links, while dotted lines show weak, transient links. It is of key importance that the figure is a snapshot only, and the position of creative elements in real-world, dynamic networks will change to a similarly integrative position elsewhere in the network at the very next moment. This is illustrated by the jump of the creative element from position A to A'.

## Creative elements have an integrated property-set

After completing the description of creative elements in a large variety of evolving networks here I show that the properties of creative elements require and predict each other and, therefore, make an integrated set of assumptions.

- **Autonomy and transient links**: creative elements are the least specialized, and are the best among all network elements to conduct an individual, autonomous life independent from the rest of the network – this independence explains why they might, and should, continuously rearrange their contacts.
- **Transient links and structural holes**: creative elements must connect elements which are not directly connected to each other. If creative elements introduced their new and unexpected content to multiple sites of a densely connected region, they would make an extremely large cumulative disorder, which would be either intolerable or would lead to a permanent change instead of a transient change. Due to the same reason creative elements must connect to hubs to allow either the dismissal, or the fast dissipation of their novel content.
- **Structural holes and network integration**: if an element connects distant modules (with transient, weak links leading to the generation of important positions of the modules involved), this element performs a continuous sampling of key information of the entire network, and, therefore has a central and integrating role in network function.
- **Network integration and creativity**: if an element is accommodating key and representative information of a whole network, it (a) may easily invent novel means to dissipate an unexpected, novel perturbation or (b) may connect distant elements of the network with ease and elegance helping them to combine their existing knowledge to cope with the novel situation. The re-formulation of the original problem (by translating it from one distant element to another), the generation of novel associations and novel solutions, flexibility, divergence and originality are all well-known hallmarks of creativity.

Creative elements are the luxury of a network operating in 'business as usual' situations. Therefore, the number of creative elements is usually very small. This situation may be characteristic of most man-made networks, such as the internet, traffic networks or power-grids. However, creative elements are the 'life insurance' of complex systems helping their survival during any unexpected damage. Therefore, the number and importance of creative elements should increase if the complex organism experiences a fluctuating environment [35,41]. The adaptation of a large group of competing organisms to fluctuating environments can be described by the process of evolution. The capacity of an organism to generate heritable phenotypic variation is called evolvability [54]. Evolvability is a selectable trait, which assumes that it is modulated by specific mechanisms [15]. Summarizing the ideas above I propose that creative elements play a crucial role in the development, inheritance and regulation of evolvability. Based on all the properties of creative elements outlined so far several methods can be designed to test or refute their existence (see Box 2).

> **Box 2. Possible proofs of principle**
> In this section I suggest several methods to test or refute the existence of creative elements. Since the properties of creative elements are linked to each other, the ideas below not only describe individual tests, but also offer a system for cross-checks, cross-correlations and for the sequential selection of creative elements [35].
> **Autonomy**
> - The internal properties of creative elements differ from the rest of the network (the number of dimensions, degrees of freedom, of this difference increase with the complexity of the element);
> - creative elements mostly act as sources in directed networks.
>
> **Network topology**
> - Creative elements are preferentially connected to hubs with large centrality;
> - creative elements provide short-cuts (decrease maximal shortest paths);
> - creative elements are in the overlaps of multiple modules maintaining roughly equal contact(s) with all modules;
> - insertion of a creative element to a network structure is predominantly occurring at positions, where it induces a large decrease in the structural entropy of the network (in other words: usually – but not always – the position of creative elements is the least random position of all network elements, giving them the most information content).
>
> **Network dynamics**
> - The internal structure of creative elements is flexible (the flexibility increases as the complexity of the element grows);
> - creative elements have more weak links than the average of the network (similarly to date hubs [40,42], they have a small number of links at a given time);
> - creative elements may be found among those elements, which have a large dynamical importance as defined by Restrepo et al. [58]
> - the behaviour of creative elements is the least predictable, if compared to the predictability of other network elements (this is also related to their extremely large autonomy).
>
> **Crisis management**
> - Creative elements have a maximal influence on the development and maintenance of cooperation in the network by mediating the conflicts of network elements and modules (as a similar finding we recently proved that innovativity helps cooperation in spatial games of social conflict [59]);
> - the number and importance (centrality) of creative elements transiently increase, if the network experiences unexpected situations regulating the evolvability of the system.

## Concluding remarks

In conclusion, I have provided in this article a description of creative elements as the network representation of active centres, and showed that their properties are consistent (1) with each other, (2) with our common knowledge on creativity and (3) with their suggested role to invent novel solutions, which integrates the knowledge of the whole network in response to unexpected situations. In addition, I would like to note that in most networks the status of the creative element is, by itself, transient. Creative elements may well be transformed into task-distributing-party-hubs [40,42] or bridges which preferentially connect two modules with strong links, or into problem-solving, specialized elements. These transformations of creative elements usually happen after repeated stress showing that the network 'learned' the novel response by re-organizing its topology, and provides the first unusual, creative solution in a regular, reliable and highly efficient manner [55]. This 'commercialization of creativity' may explain why signalling networks have isoforms of their critical nodes [48], which may replace each other in a redundant fashion, when one became engaged continuously with a specific task.

Creative elements add random elements to network behaviour inducing an increase of noise. This is highly beneficial to a certain extent, but becomes intolerable, if it exceeds a certain threshold. This threshold is high, if the hosting network lives an individual life and often meets unexpected situations. However, the same threshold becomes low, if the hosting network is part of a higher level organization, which provides a stable environment. As examples of this: 'creative cells' (e.g. ones that occur after malignant transformation) may significantly disturb the regular functions of the hosting network and finally cause its disintegration, then death; symbiotic organisms which became engulfed by another shed off a large section of their network variability [56], probably including most of their creative elements.

In the text so far I have only mentioned creative elements. Here, at the end I would like to note that the same concept may apply to links as well. I hope that the description of creative elements and links will stimulate the long range inter-modular connections [57] within, and the creative links between many brains, and will prompt further discussion and work in the field. Creativity, if combined with efficient learning, information processing and perseverance, leads to giftedness and talent – features we certainly need to understand the evolution of complexity in the network context.

**Acknowledgements**
The author would like to thank the Editor, anonymous referees and members of the LINK-Group (www.linkgroup.hu) for their helpful comments, especially Eszter Hazai and Zsolt Bikádi for the data of Figure 1. Work in the author's laboratory was supported by the EU (FP6-518230) and the Hungarian National Science Foundation (OTKA K69105).

**Glossary**

**Active centre:** In the traditional sense we call a protein segment an active centre, if it plays a key role in the catalytic action of the enzyme function displayed by the respective protein. In the network sense an active centre of the protein contains a cluster of amino acids, which have a high centrality in the amino acid network of the hosting protein. With the exception of the Introduction, throughout this paper I use the term active centre in this novel, network sense.

**Centrality**: centrality of a network element or link defines the relative importance of the element or link within the network. There are various measures of centrality using the local network topology, global properties of the whole network, or both.

**Degree**: The number of links of a given element in a network. The degree distribution is an important property of the network showing whether the network is a random network (with binomial degree distribution), a scale-free network (with a power law degree distribution), or different from both.

**Elastic network model**: In the elastic network model the protein is modelled as an ensemble of its amino acids occupying the equilibrium positions of the alpha carbon atoms, as found in the experimental three-dimensional structure. Each amino acid interacts with its neighbours, as specified by a cut-off distance, and is modelled as an oscillator linked by springs to all interacting amino acids. The model calculates the most probable modes of oscillation of each amino acid.

**Element**: The element is a single building block of a network. The element is also called a vertex in graph theory, a node or a site in physics, or an actor in sociology. Most of the times the element itself is a complex network again, like the elements of protein–protein interactions networks, the individual protein molecules can be perceived as networks of their constituting amino acids or atoms.

**Fractal**: Fractal objects are generated by a recursive process, where self-similar objects of different size are repeated and repeated again. In nature we are often talking about fractal-like behaviour, where the extent of self-similarity is not complete as in pure (and many times extremely beautiful) mathematical fractals.

**Hierarchical network**: A hierarchical organization arises in a network, when an element has a 'parent' and this 'parent' also has a 'grandparent', like in a family tree. Network hierarchy may arise at the level of modules, which may be considered as elements of a higher level network. Modules of this higher level network may be considered again as elements of an even higher level, etc.

**Hot spots**: amino acid residues of protein binding surfaces having an extraordinarily high level of binding free energy.

**Hub**: A hub is a highly connected element of the network. Usually a hub has more than 1% of total interactions.

**Module**: Modules are groups of network elements that are relatively isolated from the rest of the network, and where the elements inside the module are functionally linked to each other and have denser contacts with each other than the group with outside groups. Modules are also called network communities.

**Network topology**: The topology of a network is the precise description of the links between the elements of the network. Many special topologies are discriminated by their degree distribution.

**Scale-free topology**: Scale-free topology denotes a degree distribution of network elements that follows a power law (algebraic decay instead of, for example, exponential). These networks have many nodes with a low degree (i.e. few connected links); however, they also have a non-zero number of hubs (i.e. nodes with an unusually high number of links).

**Small world:** We call a network a small-world, when the average number of steps we need to reach any element of the network from any other elements grows only logarithmically with the number of elements in the network. In smaller networks this means that the elements are less than six steps apart from each other.

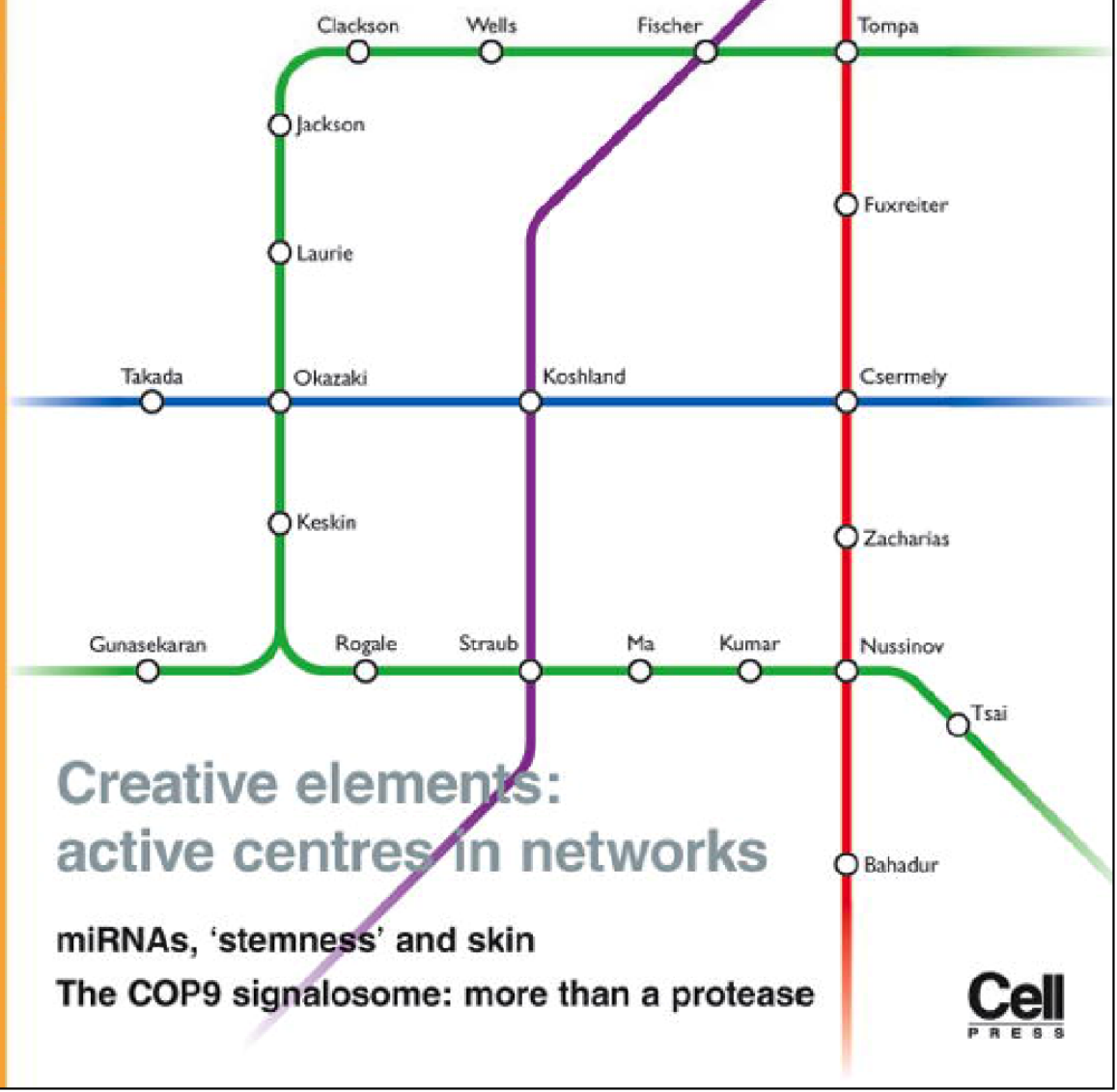